\documentclass[12pt]{article}
\usepackage{amssymb} 
\usepackage{epsfig}
\usepackage{float}
\newcommand{\be}{\begin{equation}}
\newcommand{\ee}{\end{equation}}
\newcommand{\bea}{\begin{eqnarray}}
\newcommand{\eea}{\end{eqnarray}}

\begin{document} 

\begin{center}
\bf{NEUTRINOLESS DOUBLE $\beta$-DECAY}
\footnote{A report at the Workshop in Particle Physics `` Rencontres de Physique de La Vallee d'Aoste'', La Thuile, Aosta Valley, February 29-March 6, 2004}
\end{center}

\begin{center}
S. M. Bilenky 
\end{center}
\vspace{0.1cm} 
\begin{center}
{\em  Joint Institute
for Nuclear Research, Dubna, R-141980, Russia\\}
\end{center}
\begin{center}
{\em 
SISSA, via Beirut 2-4, Trieste, 34014, Italy\\}
\end{center}
      
\begin{abstract}
The neutrinoless  double $\beta$-decay is reviewed. Model independent evidence in favor of
neutrino masses and mixing is briefly summarized. The data of the recent experiments on the 
search for $0\nu\beta\beta$-decay are presented and some future experiments are discussed.
The possible values of the effective Majorana mass, which can be predicted on the basis of the 
 neutrino oscillation data under different assumptions on the pattern of the neutrino mass spectrum, are considered. A possible model independent test of the nuclear matrix element calculations
is discussed.

\end{abstract}

\section{Introduction}

The status of the problem of the neutrino mixing drastically changed during  the last 5-6 years:
in the atmospheric Super-Kamiokande experiment \cite{SK},  the solar SNO experiment \cite{SNO,SNOsalt}
and the reactor KamLAND experiment \cite{Kamland}  strong model independent 
evidence of 
neutrino oscillations was obtained.

There are many open problems of neutrino mixing. The value of the parameter $\sin^{2}\theta_{13}$ is the most urgent one. The value of this parameter 
is crucial for the search of such fundamental effects of the tree-neutrino mixing  as CP
violation in the lepton sector. The value of the parameter $\sin^{2}\theta_{13}$ will be measured in the future reactor \cite{reactor13} and long baseline accelerator experiments 
\cite{accelerator13}.

One of the most important problem of the neutrino mixing
is the problem of the nature of the neutrinos with definite masses $\nu_{i}$: are they truly neutral Majorana particles or Dirac particles, possessing conserved total lepton number? 
The solution of this problem will have 
a profound impact on the understanding of the origin of small neutrino masses and 
neutrino mixing.

The study of the neutrinoless double $\beta$-decay ($0\nu\beta\beta$-decay)  
$$(A,Z) \to (A,Z+2) +e^{-}+ e^{-}$$
is the most sensitive method of the investigation of the Majorana nature of neutrinos with 
definite mass. We will discuss here this process.
In the next section model independent evidence of neutrino oscillations will be briefly 
reviewed. Then the data of the recent  $0\nu\beta\beta$-decay experiments  
and projects of the future experiments
will be discussed. We will consider the possible values of the effective Majorana mass, which
can be predicted from the results of the analysis of the neutrino oscillation data. 
In the last section we will discuss 
the problem of nuclear matrix elements of the  $0\nu\beta\beta$-decay. A possible test of 
the models of the nuclear matrix element calculations will be proposed.

\section{ Model independent evidence of neutrino oscillations}

Compelling model independent evidence of neutrino masses and mixing
was obtained in the atmospheric Super-Kamiokande \cite {SK}, solar SNO
\cite{SNO,SNOsalt} and reactor KamLAND \cite{Kamland} 
experiments.

In the Super-Kamiokande (SK) experiment zenith angle $\theta_{z}$ 
dependence of the electron and muon atmospheric neutrino events 
was studied in details.
If there are no neutrino oscillations the number of the high-energy neutrino-induced 
$\mu (e) $ events 
$N_{\mu , e}(\cos \theta_{z}) $ must satisfy the
symmetry relation
$$
N_{\mu , e} (\cos \theta_{z}) = N_{\mu , e}(-\cos \theta_{z}).
$$
For the high-energy muon neutrinos this relation is strongly violated:
the significant deficit of the up-going muons is observed. For the ratio of the total numbers of the up-going and down-going muon neutrinos  
the following value
\be
\left(\frac{U}{D}\right)_{\mu}= 0.54 \pm 0.04 \pm 0.01
\label{1}
\ee
was obtained. Here $U$ is the total number of the up-going muons 
($-1 \leq \cos \theta_{z} \leq -0.2$) and $D$ is the total number of the down-going 
muons 
($0.2 \leq \cos \theta_{z} \leq 1$).

The up-going muons are produced by neutrinos passing distances
from about $500\,\rm{km}$ to  about
$13000 \, \rm{km}$ and  
the down-going muons are produced by neutrinos traveling distances from 
about $20\, \rm{km}$
to about $500\, \rm{km}$. The ratio (\ref{1}) clearly demonstrates the dependence of 
the flux of the muon neutrinos on the distance between the neutrino production region in the atmosphere and   neutrino detector.

The data of the SK experiment and other atmospheric neutrino experiments 
MACRO \cite{Macro} and SOUDAN 2 \cite{Soudan} are described by 
the two-neutrino $\nu_{\mu}\to \nu_{\tau}$  oscillations. From the two-neutrino analysis of the SK data 
the following best-fit values of the oscillation parameters were found
\be
(\Delta m^{2})_{\rm{SK}}=2\cdot 10^{-3}\rm{ eV}^{2};\,
(\sin^{2}2 \theta)_{\rm{SK}}=1.0 \,~
(\chi^{2}_{\rm{min}}= 170.8/ 170\,\rm{d.o.f.}) 
\label{2} 
\ee
At the  90\% CL the oscillation parameters are in the ranges
\be
1.3 \cdot 10 ^{-3}\leq (\Delta m^{2})_{\rm{SK}} \leq 3.0 
\cdot 10 ^{-3}\rm{eV}^{2}; \,~~(\sin^{2}2 \theta)_{\rm{SK}}>0.9 
\label{3}
\ee
In the SNO experiment \cite{SNO,SNOsalt} solar neutrinos from the decay
$^8 \rm{B} \to ^8 \rm{Be} +e^{+} + \nu_{e}$ 
are detected via the observation 
of the CC reaction
\be
\nu_{e}+d\to e^{-} +p+p
\label{4}
\ee
and the NC reaction

\be
\nu_{l}+d\to \nu_{l}+n+p\,~~ (l =e, \mu,\tau)
\label{5}
\ee
For the total flux of $\nu_{e}$ detected via CC reaction (\ref{4})
the following value
\be
\Phi_{\nu_{e}}^{\rm{SNO}}=
({1.59^{+0.09}_{-0.07}\mbox{(stat.)}^{+0.06}_{-0.08}~\mbox{(syst.)}} ) \cdot 10^{6}\,~
cm^{-2}s^{-1} 
\label{6}
\ee
was obtained in the SNO experiment.

For the total flux of 
$\nu_{e}$, $\nu_{\mu}$ and $\nu_{\tau}$ detected via NC reaction (\ref{5})
the significantly larger value
\be
\sum_{l=e,\mu,\tau}\Phi_{\nu_{l}}^{\rm{SNO}}=(5.21\pm 0.27\pm 0.38) \cdot 10^{6}\,~
cm^{-2}s^{-1} 
\label{7}
\ee
was found.
Eq.(\ref{6}) and Eq.(\ref{7}) give us
a model independent evidence of the transitions 
of the original solar $\nu_{e}$ into  $\nu_{\mu,\tau}$.

The results of all solar neutrino experiments \cite{SNO,SNOsalt,Cl,GNO,SAGE,SKsol}
 are described by the two-neutrino $\nu_{e}\to\nu_{e} $ 
survival probability in matter.
From the analysis of the data in the most preferable LMA region 
the following best-fit values of the 
oscillation parameters were found \cite{SNO}
\be
(\Delta m^{2})_{\rm{sol}}=5\cdot 10^{-5}\rm{eV}^{2};
\,(\tan^{2}\theta)_{\rm{sol}}=0.34;\,~
\chi^{2}_{\rm{min}}= 57/72\, \rm{d.o.f.}.
\label{8}
\ee

In the KamLAND experiment \cite{Kamland}
electron antineutrinos from many reactors in Japan and Korea are detected via observation of 
the process
$$\bar\nu_{e}+p \to e^{+}+n $$

The average distance  between reactors and the detector in this experiment is 180 km.
For
the ratio of the total numbers of the observed and expected events  
the following value 
$$
\frac{N_{obs}}{N_{exp}} = 0.611 \pm 0.085 \pm 0.041$$
was obtained. This result 
is a clear 
evidence of the disappearance of $\bar\nu_{e}$'s on the way from the reactors to the detector.

The KamLAND data are described by the two-neutrino
$\bar\nu_{e}\to\bar \nu_{e} $ oscillations in vacuum. 
The best-fit values of the oscillation parameters which were found
from the 
analysis of the data 
\be
(\Delta m^{2})_{\rm{KL}}= 6.9 \cdot 10^{-5}\rm{eV}^{2};\,~
(\sin^{2}2\,\theta) _{\rm{KL}}=1.
\label{9}
\ee
are compatible with the solar LMA values.

From common analysis of the solar and KamLAND data (assuming CPT)
the following best-fit values of the parameters
\be
(\Delta m^{2})_{\rm{sol+KL}}=7.1\cdot 10^{-5}\rm{eV}^{2};
\,(\tan^{2}\theta)_{\rm{sol+KL}}=0.41
\label{10}
\ee
were obtained \cite{SNOsalt}.

In \cite{Sand} from the global two-neutrino analysis of the solar and the KamLAND data the following 
90\% CL ranges of the parameters were found
\be
6.0\cdot 10^{-5}\leq  (\Delta m^{2})_{\rm{sol+KL}}\leq 8.7\cdot 10^{-5}\rm{eV}^{2};\,~
 0.25 \leq (\sin^{2}\theta)_{\rm{sol+KL}}\leq 0.37.
\label{11}
\ee
SK atmospheric neutrino evidence of neutrino oscillations was 
confirmed by the long baseline accelerator K2K experiment \cite{K2K}.
The distance between neutrino source (KEK accelerator) and SK detector is about
250 km.
Average $\nu_{\mu}$ energy is 1.3 Gev. 
In the K2K experiment
56 $\nu_{\mu}$  events were observed. The expected number of the events 
is equal to $80.1 ^{+6.2}_{-5.4}$.
The best-fit values 
of the two oscillation parameters found from
the two-neutrino analysis of the data 
$$
(\sin^{2}2\,\theta)_{\rm{K2K}}=1;\,~ (\Delta m^{2})_{\rm{K2K}}= 2.8\,~10^{-3}\,~ \rm{eV}^{2}
$$
are compatible with the atmospheric values (\ref{3}).

The negative 
results of the CHOOZ \cite{CHOOZ} and Palo Verde \cite{PVerde} reactor experiments 
are very important for the neutrino
mixing. In these experiments  reactor $\bar \nu_{e}$ 's are detected via the observation of the 
classical process 
$\bar\nu_{e}+p \to e^{+}+n $. The distances between reactors and detectors 
in these experiments were 
about 1 km. No disappearance of $\bar\nu_{e}$'s were observed.
In the CHOOZ 
experiment for the ratio of the total numbers of observed and expected
events the following value
$$
\frac{N_{obs}}{N_{exp}} = 1.01 \pm 2.4\% \pm 2.7 \%
$$
was found.
From the CHOOZ two-neutrino exclusion curve at 
$\Delta m^{2}= 2 \cdot 10^{-3}\rm{eV}^{2}$
(the SK best-fit value) the following upper bound
\be
(\sin^{2}2\theta)_{\rm{CZ}} \leq 2\cdot 10^{-1}
\label{12}
\ee
can be obtained.

\section{Neutrino oscillations in the framework of three-neutrino mixing}

Neutrino oscillation data are analyzed under three basic assumptions (see, for example, 
 \cite{BGG}):
\begin{itemize}
\item
Three flavor neutrinos $\nu_{e}$, $\nu_{\mu}$ ,$\nu_{\tau}$ ( and antineutrinos 
$\bar\nu_{e}$, $\bar\nu_{\mu}$ ,$\bar\nu_{\tau}$ ) exist in nature.

\item
The Lagrangian of the interaction of the flavor neutrinos
is given by the Standard Model. The
lepton charged current and neutrino neutral current have the form

\be
j^{\mathrm{CC}}_{\alpha} = 2\,\sum_{l=e,\mu,\tau} \bar \nu_{lL} \gamma_{\alpha}l_{L};\,~~
j^{\mathrm{NC}}_{\alpha} =\sum_{l=e,\mu,\tau} \bar \nu_{lL}\gamma_{\alpha}\nu_{lL}.
\label{13}
\ee

\item
Neutrino mixing
\be
\nu_{{l}L} = \sum^{3}_{i=1}U_{{l}i} \, \nu_{iL} 
\label{14}
\ee
takes place. Here  $\nu_{i}$ is the field on neutrino with mass $m_{i}$
and $U$ is the unitary PMNS \cite{BP,MNS} mixing matrix.

\end{itemize}
The three-neutrino scheme (\ref{14})
is the minimal scheme of neutrino mixing:
the number of the massive neutrinos $\nu_{i}$ is equal the the number 
of the flavor neutrinos 
$\nu_{l}$ (three). If the number of light neutrinos $\nu_{i}$ is larger than three, sterile neutrinos must exist. An indication in favor of neutrino oscillations 
with the third relatively large neutrino mass-squared 
difference ($\simeq 1 \,\rm{eV}^{2}$) was obtained in the single accelerator 
LSND experiment \cite{LSND}. The  LSND result 
needs confirmation. It will be checked 
by the 
MiniBooNE experiment at the Fermilab
\cite{MiniB} .

Neutrino oscillation data are compatible with two types of neutrino
mass spectra\footnote{$\Delta m^{2}_{ki}=m^{2}_{k}-
m^{2}_{i}$. 
Notice that different notations for neutrino masses are used for NS and IS spectra}.
\begin{enumerate}
\item Normal  mass spectrum (NS):

$$m_{1}< m_{2}< m_{3};\,~~
\Delta m^{2}_{21} \simeq\Delta m^{2}_{\rm{sol-KL}};\,~
\Delta m^{2}_{32} \simeq\Delta m^{2}_{\rm{SK}}.$$

\item
Inverted mass spectrum (IS):

$$m_{3}< m_{1}< m_{2};\,~~
\Delta m^{2}_{21} \simeq\Delta m^{2}_{\rm{sol-KL}};\,~
\Delta m^{2}_{31} \simeq -\Delta m^{2}_{\rm{SK}}$$

\end{enumerate}

Here $\Delta m^{2}_{\rm{SK}}$ and $\Delta m^{2}_{\rm{sol-KL}}$ 
are neutrino mass-squared differences, which are determined 
from the two-neutrino analysis
of the atmospheric SK and the solar-KamLAND data
(see (\ref{2}), (\ref{3}) ,  (\ref{10}) and (\ref{11}) ).

From neutrino oscillation data it follows that two independent neutrino mass-squared differences satisfy the inequality:
\be
 \Delta m^{2}_{21}\ll \Delta m^{2}_{32}.
\label{15}
\ee
In the framework of the three-neutrino mixing we have (see \cite{BGG})
\be
4\, |U_{e3}|^{2}\, (1- |U_{e3}|^{2} )= (\sin^{2}2\theta)_{\rm{CZ}}.
\label{16}
\ee
If we take into account solar data, from (\ref{12}) and (\ref{16}) we obtain 
that 
\be
|U_{e3}|^{2}\leq 5\cdot 10^{-2} 
\label{17}
\ee

{\em The present-day picture of neutrino oscillations} is
determined by the inequalities (\ref{15}) and (\ref{17}) (see, for example, \cite{BGG}).
In the leading approximation  neutrino oscillations in the experiments with 
 $L/E\simeq 10^{3}$
($L$ is the source-detector distance in m (km), $E$ is  the neutrino energy in MeV (GeV))
are two-neutrino $\nu_{\mu} \to \nu_{\tau}$ oscillations.
Solar neutrino transitions in matter (and antineutrino oscillations in reactor experiments with
$L/E\simeq 10^{5}$) are
$\nu_{e}\to \nu_{\mu,\tau}$ ($\bar\nu_{e}\to  \bar\nu_{\mu,\tau}$) oscillations.
Solar $\nu_{e}$ and reactor $\bar\nu_{e}$ survival probabilities depends on two parameters 
$\Delta m^{2}_{21}$ and $\sin^{2}\theta_{12}$
(if CPT is assumed) and are given 
by the standard two-neutrino expressions in matter and in vacuum, respectively.

The observation of neutrino oscillations means
that the flavor lepton numbers $L_{e}$, $L_{\mu}$ and $L_{\tau}$
are not conserved by a neutrino mass term. 
There exist two theoretical possibilities in this case
\begin{enumerate}
\item
The total lepton number
$L = L_{e} + L_{\mu} +L_{\tau}$ is conserved. In this case fields of neutrinos with definite masses 
$\nu_{i}(x)$ are {\em Dirac fields} of 
neutrinos (L=1) and antineutrinos (L=-1).
\item
The total lepton number $L$ is violated. In this case the fields of
neutrinos with definite masses $\nu_{i}(x)$ are {\em Majorana fields} of neutrinos
(with antineutrinos identical to neutrinos). The fields $\nu_{i}(x)$ satisfy the Majorana condition 
\be
\nu^{c}_{i}(x)= C\,\bar\nu^{T}_{i}(x)=\nu_{i}(x),
\label{18}
\ee
where $C$ is the charge conjugation matrix.
\end{enumerate}

After the discovery of the neutrino mixing
the problem of
the nature of massive neutrinos (Dirac or Majorana?) is
one of the most
fundamental. There is no doubt that the solution of this problem will have very important
impact on our 
understanding of the origin of small neutrino masses and neutrino mixing.

The investigation of the neutrino oscillations is an extremely sensitive method of 
the measurement of the very small neutrino mass-squared differences \cite{BilPont}.  
This method does not allow, however,  to reveal the
nature
of massive neutrinos  \cite{BHP}. 

In fact mixing matrices in the Majorana and Dirac cases are connected by the relation
$$
U^{Mj}= U^{D}\,~S(\beta),
$$
where $S_{ik}(\beta)=e^{i\,\beta_{k}}\,\delta_{ik} $ is a diagonal phase matrix. The matrix
$S(\beta) $ does not give contribution to the transition probability
$$
{\mathrm P}(\nu_{l} \to \nu_{l'}) =|\,\sum_{i}U_{l'  i} 
\,~ e^{- i \Delta m^2_{i1} \frac {L} {2E}} \,~U_{l i}^*\, |^2 .
$$
Thus, we have
\be
{\mathrm P}^{Mj}(\nu_{\alpha} \to \nu_{\alpha'}) =
{\mathrm P}^{D}(\nu_{\alpha} \to \nu_{\alpha'})
\label{19}
\ee
The nature of the massive neutrinos can be revealed only via the investigation of the
processes in which the total lepton number 
$L$ is not conserved. The most sensitive 
to small Majorana 
neutrino masses process is
neutrinoless double $\beta $- decay of even-even nuclei
(see reviews \cite{Doi,BPet,Faessler,Suhonen,Vergados,Elliot,BGGM}):

$$
(\rm{A,Z}) \to (\rm{A,Z +2}) +e^- +e^-.
$$

In the case of the small Majorana neutrino masses the half-life of this process is given by the following general expression

\be
\frac{1}{T^{0\,\nu}_{1/2}(A,Z)}=
|m_{ee}|^{2}\,|M^{0\,\nu}(A,Z)|^{2}\,G^{0\,\nu}(E_{0},Z).
\label{20}
\ee
Here 
$|m_{ee}|$ is 
the effective Majorana mass
\be
|m_{ee}| = |\sum_{i} U^{2}_{ei}\, m_{i}|,
\label{21}
\ee
$M^{0\,\nu}(A,Z) $ is nuclear matrix element, which 
depends only on nuclear properties, and 
$G^{0\,\nu}(E_{0},Z)$ is known phase-space factor
($E_{0}$ is the energy release).
In the next sections we will discuss
\begin{enumerate}
\item
Existing data and future experiments.
\item

Possible values of the effective Majorana mass $|m_{ee}|$.

\item
The problem of nuclear matrix elements.
\end{enumerate}

\section{Existing $0\nu\beta\beta$-decay data and future experiments}

Neutrinoless double $\beta$ decay is allowed for such even-even nuclei
for which usual  $\beta$ decay is forbidden by the conservation of energy. 
 There are several nuclei of this type:
$$
^{76} \rm{Ge} (2.039),\, ^{130} \rm{Te} (2.528), \,
^{136} \rm{Xe} (2.480),  ^{100} \rm{Mo} (3.034 ),
^{150} \rm{Nd} (3.367 )
$$
and others. In the brackets the energy release $E_{0}$ in MeV is given. 
This is an important characteristic 
of the $0 \nu \beta \beta$-decay: the decay probability is proportional to 
$E^{5}_{0}$.

The results of many experiments on the search for $0 \nu \beta \beta$ -decay
are available at present (see \cite{Zdesenko,Gratta,PDG}).        
The most stringent  lower bounds on the half-life 
of $0\,\nu \beta\,\beta $-decay 
was reached in the Heidelberg-Moscow \cite{HM} and CUORICINO \cite{Cuoricino} 
experiments.
 
The detector (and source) of the Heidelberg-Moscow  experiment \cite{HM}
consists of 5 crystals of 86\%  enreached
$^{76} \rm{Ge}$ of the total mass about 11kg .
For the half-life the lower bound
\be
T^{0\nu}_{1/2}\geq 1.9 \cdot 10^{25}\, y\,~~ (90\% \,\rm{CL})
\label{22}
\ee
has been found. Taking into account different calculations of the nuclear matrix element,
from (\ref{22}) for the effective Majorana mass $|m_{ee}|$ 
the following upper bounds
\be
|m_{ee}| \leq (0.3-1.2)\,~\rm{eV}\,. 
\label{23}
\ee
were  obtained.

In the cryogenic experiment CUORICINO \cite{Cuoricino}  
$\rm{Te}\rm{O}_{2} $
crystals  with a total mass 40.7 kg are employed.
 For the half-life of 
$^{130} \rm{Te}$ in this experiment the following lower bound
\be
T^{0\nu}_{1/2}\geq 7.5 \cdot 10^{23}\, \rm{years}
\label{24}
\ee
was reached recently.
From (\ref{24}) for the effective Majorana mass
the upper bounds 
\be
|m_{ee}| \leq (0.3-1.7)\,~\rm{eV}
\label{25}
\ee
were obtained.

Many projects of new experiments on the search for the neutrinoless 
double $\beta$-decay of different nuclei 
are under research and development at present (see \cite{Elliot,Gratta,Avignone} ).

The main goal of the future experiments is to reach the sensitivity
$|m_{ee}| \simeq \rm{a \,~few} 10^{-2}\rm{eV}$.
This goal can be accomplished
 by  detectors with mass about 1 ton or more, which have  a good energy resolution,
 low background and an efficient signature for $0\nu \beta \beta $ events.

The  experiment CUORE \cite{Cuoricino} will be a continuation of the  
CUORICINO experiment.
Cryogenic detector 
will consist of 1000 $ \rm{Te\,O_{2}}$ crystals
operated at a temperature 10 mK. The total mass of the detector will be about 800 kg. The
expected  resolution at $E_{0}= 2.528$ MeV is 5 keV.
For the half-life of $^{130} \rm{Te}$
the value
\be
T^{0\nu}_{1/2}\simeq 9.5\cdot 10^{26}\, \rm{years}
\label{26}
\ee
is envisaged. This corresponds to the sensitivity
\be
|m_{ee}| \simeq (2-5.2)\cdot 10^{-2}\,~\rm{eV}
\label{27}
\ee
In the EXO experiment \cite{Gratta} up to $ 10$ tons of
60-80 \% enreached $^{136} \rm{Xe}$ are planned to use. 
An important feature of this experiment  
is a laser tagging of
$ \rm{Ba}^{+}$ ions, produced in
the recombination of $\rm{Ba}^{++}$ ions from the 
decay 
$^{136} \rm{Xe}\to ^{136} \rm{Ba}^{++}+e^{-}+e^{-}$.
The detection of $ \rm{Ba}^{+}$ ions will provide large background reduction. 
The value
\be
T^{0\nu}_{1/2}\simeq 1\cdot 10^{28}\, \rm{years}
\label{28}
\ee
is expected.  It corresponds to the sensitivity
\be
|m_{ee}| \simeq (1.3-3.7)\cdot 10^{-2}\,~\rm{eV}
\label{29}
\ee

The GENIUS experiment \cite{Klap}
 will be a development of the Heidelberg-Moscow experiment. 
 About 1 ton of  86 \% enreached $^{76} \rm{Ge}$  will be embadded in a large 
liquid nitrogen cryostat .
The liquid nitrogen will provide effective  shielding from the external background.
For the half-life a value
\be
T^{0\nu}_{1/2}\simeq 1\cdot 10^{28}\, \rm{years}
\label{30}
\ee
is expected. This value corresponds to the sensitivity
\be
|m_{ee}| \simeq (1.3-5.0)\cdot10^{-2}\,~\rm{eV}
\label{31}
\ee

In the MAJORANA experiment \cite{Avignone}, which will be the continuation of the IGEX
experiment \cite{IGEX}, about 500  kg of   86 \% enreached $^{76} \rm{Ge}$ will be used. 
The main background is expected 
from the decay   $^{68} \rm{Ge}\to ^{68} \rm{Ga} +e^{+}+\nu_{e} $.
 It will be suppressed by 
the segmentation of the detector and  effective pulse shape analysis of the` signal.
In the MAJORANA experiment the value
\be
T^{0\nu}_{1/2}\simeq 4\cdot 10^{27}\, \rm{years}
\label{32}
\ee
is expected. It corresponds to the sensitivity 
\be
|m_{ee}| \simeq (2.1-7.0)\cdot 10^{-2}\,~\rm{eV}.
\label{33}
\ee

\section{Effective Majorana mass}

The effective Majorana mass is determined by the absolute values of the 
neutrino masses, mixing angles and CP phases. Let us discuss first these three ingredients.

\begin{enumerate}
\item Neutrino masses

From neutrino oscillation data only neutrino mass-squared differences can be inferred.
In the case of the NS neutrino mass spectrum the neutrino masses are given by the relations
$$m_{2} \simeq \sqrt{m^{2}_{1}+\Delta m^{2}_{21}};\,~
m_{3} \simeq \sqrt{m^{2}_{1}+\Delta m^{2}_{32}}$$
For the IS neutrino mass spectrum we have
$$m_{1} \simeq m_{2}\simeq \sqrt{m^{2}_{3}+|\Delta m^{2}_{31}}|$$

From existing data only 
upper bounds of the lightest mass $m_{\rm{min}} (m_{1}\,~\rm{or}\,~ m_{3})$
can be obtained.
From the data of the Troitsk \cite{Troitsk} and Mainz \cite{Mainz} tritium 
$\beta$-decay experiments 
the following bounds were found
\be
m_{\rm{min}} \leq 2.05\,~ (2.3) \,~\rm{eV}
\label{34}
\ee
More stringent bound

$$m_{\rm{min}}  \leq 0.6\,~\rm{eV}$$
was obtained from the analysis \cite{Tegmark} of the data of WMAP and Sloan Digital 
Sky Survey Collaborations.
\item Mixing angles.

The elements $U_{ei}$ (i=1,2,3) are given by 
$$U_{ei}=|U_{ei}| \,e^{\,i\,\alpha_{i}},$$
where $\alpha_{i}$ are Majorana CP-phases. 
If we introduce the angle $\theta_{13}$
in such a way that 
$$|U_{e3}|^{2}=\sin^{2}\theta_{13}$$
from the unitarity relation 
$\sum^{3}_{i=1} |U_{ei}|^{2} =1$
we have
$$|U_{e1}|^{2}=\cos^{2}\theta_{13}\,\cos^{2}\theta_{12};\,
|U_{e2}|^{2}=\cos^{2}\theta_{13}\,\sin^{2}\theta_{12}.$$

The parameter  $\sin^{2}\theta_{12}$ can be determined from the results of the solar and 
KamLAND experiments. From the analysis of the existing data 
the best fit value (\ref{10}) and a range (\ref{11}) 
were found.  

Only upper bound of the parameter $\sin^{2}\theta_{13}$ is known today. From the exclusion curve obtained in the CHOOZ experiment the bound (\ref{17}) was obtained. 

\item Majorana phases 

Majorana phases $\alpha_{i}$ are unknown. In the case of the CP 
invariance in the lepton sector the elements of the mixing matrix satisfies the relation
\cite{BNP}:

\be
U_{ei}=U^*_{ei} \,\eta_{i} ,  
\label{35}
\ee
where $\eta_{i}= i\,\rho_{i}\,~~\rho_{i}=\pm 1$ 
is the CP-parity of the Majorana neutrino $\nu_{i}$.
Thus, in the case of the CP invariance
$$U^{2}_{ei}=i\,|U_{ei}| ^{2}\,\rho_{i} $$  
and the effective Majorana mass takes the form
\be
|m_{ee}| = |\sum^{3}_{i=1}| U_{ei}|^{2}\,\rho_{i}\, m_{i}|
\label{36}
\ee
From this equation it follows that in the case of the different CP parities of 
Majorana neutrinos with definite masses cancellations of their contributions to the effective Majorana mass 
can take place.
In the general case of the CP violation $|m_{ee}| $ depends on two Majorana phase differences.

\end{enumerate}

The value of the effective Majorana mass $|m_{ee}| $ strongly depends on the pattern of 
the neutrino mass spectrum and lightest neutrino mass \cite{BGGKP}.
Three types of the neutrino mass spectrum are 
usually considered.  
\begin{itemize}

\item
Neutrino mass hierarchy \footnote{Notice that masses of quarks (up and down) and
charged leptons follow the hierarchy of the type (\ref{37}).}

\be
m_{1}\ll m_{2}\ll m_{3}.
\label{37}
\ee
In the case of the neutrino mass hierarchy $m_{2}$ and $m_{3}$ are determined by the ``solar-KamLAND''  
and ``atmospheric''
mass-squared differences
\be
m_{2}\simeq \sqrt{\Delta m^{2}_{21}};\,~~
m_{3}\simeq \sqrt{\Delta m^{2}_{32}}
\label{38}
\ee
and the lightest neutrino mass $m_{1}$ is small: 
$m_{1}\ll\sqrt{\Delta m^{2}_{21}}\simeq 8.4\cdot 10^{-3}\rm{eV}$.

Neglecting the contribution of $m_{1}$ to the effective Majorana neutrino mass, we have
\be
|m_{ee}|\simeq\left |\,
\cos^{2} \theta_{13}\, \sin^{2} \theta_{12} \,
 \sqrt{\Delta m^{2}_{21}} + 
 e^{i\,2\alpha_{32}}  \sin^{2} \theta_{13}\, \sqrt{\Delta m^{2}_{32}}\,\right|,
\label{39}
\ee
where $\alpha_{32} =\alpha_{3}- \alpha_{2}$. 
It follows from (\ref{39}) that $|m_{ee}|$ is small:
the first term is small because of the smallness of $\sqrt{\Delta m^{2}_{21}}$; the 
contribution of the ``large mass''
$\sqrt{\Delta m^{2}_{32}}$ is suppressed by the smallness of the parameter $\sin^{2} \theta_{13}$.
From (\ref{3}) (\ref{11}) (\ref{17}) and (\ref{39}) we obtain the following upper bound
(90 \% CL):

\be
|m_{ee}| \leq 6.0\cdot 10^{-3}\rm{eV}
\label{40}
\ee

The bound (\ref{40}) is significantly smaller than the sensitivity of the future experiments on the search for $0\nu\beta\beta$-decay.
The observation of the neutrinoless double $\beta$-decay 
in the experiments of the next generation 
would mean that the neutrino masses do not follow the hierarchy (\ref{37}).

\item 
Inverted hierarchy of neutrino masses
\be
m_{3}\ll m_{1}< m_{2}.
\label{41}
\ee
In the case of the inverted mass hierarchy $m_{1}$ and $m_{2}$ are determined by ``atmospheric''
mass-squared difference

\be
m_{1}\simeq m_{2}\simeq \sqrt{|\Delta m^{2}_{31}|}
\label{42}
\ee
and the lightest neutrino mass  $m_{3}$ is small: 
$m_{3}\ll\sqrt{|\Delta m^{2}_{31}|}\simeq 4.5\cdot 10^{-2}\rm{eV}$.

Neglecting small contributions of $m_{3}$ and  $\sin^{2} \,\theta_{13}$, for the effective Majorana mass we have

\be
|m_{ee}|\simeq \sqrt{|\Delta m^{2}_{31}|}\, |\sum_{i=1,2}U^{2}_{ei}|
\simeq \sqrt{ |\Delta m^{2}_{31}|}\,~
(1-\sin^{2} 2\,\theta_{12}\,\sin^{2}\alpha_{21})^{\frac{1}{2}},
\label{43}
\ee
where $\alpha_{21} =\alpha_{2}-\alpha_{1}$. In the case of the CP invariance 
$\alpha_{21} = \frac{\pi}{4}(\rho_{2} - \rho_{1})=0, \pm \frac{\pi}{2}$.

From Eq.(\ref{43}) for the effective Majorana mass we have the following range 

\be
\cos 2\,\theta_{12} \,\sqrt{ |\Delta m^{2}_{31}|} \leq |m_{ee}|
\leq\sqrt{ |\Delta m^{2}_{31}|}, 
\label{44}
\ee
where the upper and lower bounds correspond to the cases of the CP conservation: 
the upper bound corresponds to the case of equal CP parities of $\nu_1$ and $\nu_2$
and the lower bound corresponds to the case of opposite CP parities.

From (\ref{44}) and (\ref{11}) at 90\% CL for the effective Majorana mass
$|m_{ee}|$ we have the range
\be
0.26\,\sqrt{ |\Delta m^{2}_{31}|} \leq |m_{ee}|
\leq\sqrt{ |\Delta m^{2}_{31}|} 
\label{45}
\ee
Thus, in the case of the inverted hierarchy 
of neutrino masses
the scale of the effective Majorana mass  is determined by
$\sqrt{|\Delta m^{2}_{31}|}$.
 From (\ref{45}) and (\ref{3}) we have
\be
0.9 \cdot 10^{-2}\rm{eV} \leq |m_{ee}|\leq
5.5 \cdot 10^{-2}\rm{eV}
\label{46}
\ee
The values of the effective Majorana mass in the range (\ref{46}) can be reached 
in
$0\nu\beta\beta$-decay experiments
of the next generation. Let us stress that in the case of the inverted hierarchy 
of the neutrino masses the value of $|m_{ee}|$ can not be  smaller than 
$\simeq 1\cdot 10^{-2}\rm{eV} $. This is connected with the fact that  
the value $\sin^{2} 2\,\theta_{12}=1$ (maximal 1-2 mixing) is excluded at 5.4 $\sigma$ level 
by the solar and KanLAND data \cite{SNOsalt}.

From Eq. (\ref{43}) we find 
\be
\sin^{2}\alpha_{21} \simeq \left( 1 -\frac{|m_{ee}|^{2}}
{ |\Delta m^{2}_{31}|}
\right)\,~\frac{1}{\sin^{2}2\,\theta_{12}}\,.
\label{47}
\ee
If the problem of the nuclear matrix elements will be solved (see the next section for discussion), 
the experiments on the measurement of the half-life 
of $0\nu\beta\beta$-decay could allow to obtain an information on
 the the value of the CP parameter $\sin^{2}\alpha_{21}$ \cite{BGKP}.

\item
Practically degenerate neutrino masses

If  the lightest neutrino mass is much larger than  
$ \sqrt{ \Delta m^{2}_{32}}\simeq 4.5\cdot 10^{-2}\rm{eV}$
in this case neutrino masses are practically degenerate:
\be
m_{1}\simeq m_{2}\simeq m_{3}.
\label{48}
\ee

For the degenerate neutrino masses the effective Majorana mass 
\be
|m_{ee}| \simeq  m_{0}\,~|\sum^{3}_{i=1}U^{2}_{ei}|  \simeq  m_{0}\,~
(1-\sin^{2} 2\,\theta_{12}\,\sin^{2}\alpha)^{\frac{1}{2}},
\label{49}
\ee
depends on two parameters: $\sin^{2}\alpha $ ($\alpha$ is the Majorana CP phase difference) 
 and a common mass $m_{0}$.
From (\ref{49}) and (\ref{11}) we have the range
\be
0.26\, ~m_{0}\leq |m_{ee}|\leq m_{0}
\label{50}
\ee
Thus, in the case of the practically degenerate Majorana neutrino mass spectrum
the scale of the effective Majorana mass $|m_{ee}|$ is determined by (unknown) common mass $m_{0}$.

The mass $m_{0}$ can be determined from the data of the   experiments  on the   measurement of the high-energy part of  the $\beta$-decay spectrum
of tritium and from cosmological data.
In the tritium experiment KATRIN \cite{Mainz,Katrin}, now at preparation, 
the sensitivity  $m_{0}\simeq 0.2\,~ \rm{eV}$ is expected.

If in the  future $0\nu\beta\beta$-decay experiments 
it will be found that the value
of the effective Majorana mass is significantly larger than
$\sqrt{\Delta m^{2}_{32}}\simeq 4.5 \cdot 10^{-2}\rm{eV}$ 
from $0\nu\beta\beta$-decay data
an information about the value of the common mass $m_{0}$ can be inferred:

\be
|m_{ee}|\leq m_{0}\leq \frac{ |m_{ee}| }{\cos 2\,\theta_{12} }
\label{51}
\ee
Using the existing data (see (\ref{11})) we have

$$|m_{ee}|\leq m_{0}\leq  3.8\,~|m_{ee}| $$.

In conclusion we would like to emphasize that the measurement of the effective Majorana mass $|m_{ee}|$ could allow to obtain an important information on 
the pattern of 
the Majorana neutrino mass spectrum.
\end{itemize}

\section{The problem of the nuclear matrix elements}
Neutrinoless double $\beta$-decay is a second order in the Fermi constant 
$G_{F}$ process
with a virtual neutrino.
For small neutrino masses (much smaller than the bounding energy of 
nucleons in nuclei)  the
matrix element of the $0\nu\beta\beta$ -decay is factorized in the form of 
a product of the effective 
Majorana mass $|m_{ee}|$, which depends on neutrino masses and elements 
$U^{2}_{ei}$ of the mixing matrix,
and nuclear matrix element (NME), which is determined only by the strong interaction. 
The NME is a matrix element of 
the chronological product of the two CC hadronic 
currents and the neutrino propagator. It can not be connected with other observables.
In the calculation of NME
many intermediate states must be taken into account.

Two basic methods of the calculations of NME are used :
quasiparticle random phase approximation (QRPA) and nuclear shell model (NSM).
Many
calculations of NME of different nuclei, based on these approximate approaches, exist in 
literature (see \cite{Faessler,Suhonen,Elliot}).
The results of different  calculations of NME differ by about factor three or more.
For example,
if we assume that $|m_{ee}|= 5\cdot 10^{-2}\,\rm{eV}$
from different calculations of NME 
for the half-life of the 
$0\nu \beta\beta $-decay of $^{76} \rm{Ge}$ the values in the  range 
$$ 6.8\cdot 10^{26} \rm{y} \leq T^{0\nu}_{1/2}(^{76} \rm{Ge})
\leq 70.8 \cdot 10^{26} \rm{y}$$
can be obtained \cite{Elliot}.

Recently a progress in the calculation of NME in the framework of 
QRPA have been achieved \cite{Rodin}.
The nuclear matrix elements of 
$0\nu \beta\beta $- decay 
of $^{76} \rm{Ge}$, $^{100} \rm{Mo}$, $^{130} \rm{Te}$ and $^{136} \rm{Xe}$
were calculated
with the values of the  
parameter  of the
effective particle-particle interaction $g_{pp}$ 
determined from the measured life-time of the $2\nu \beta\beta $- decay.
It was shown that the values of nuclear matrix elements of the
$0\nu\beta\beta$-decay of
these nuclei are stable under the change of the nuclear
potential and the number of single particle states used as a basis.
Moreover the matrix elements calculated in 
different QRPA  models 
are practically the same
(differ not more than
10\%.)
\section{Possible test of the models of NME calculations}

Taking into account all uncertainties connected with the calculations of 
the nuclear matrix elements of the $0\nu \beta\beta $- decay,
 it will be very important 
to find a possibility to test  NME calculations. We will discuss here a  
possible test, based on the factorization property of the matrix elements of the 
$0\nu \beta\beta $- decay (see\cite{BilGri}).

The proposed test can be realized  if  neutrinoless double $\beta$-decay of {\em several nuclei} 
$A_i, Z_i$ (i=1,2,...) is observed. Using a model $M$ of the calculation of the nuclear matrix elements, from Eq.(\ref{20}) the value of the parameter $|m_{ee}|^{2}_{A_i, Z_i}(M)$ 
can be determined. 
The model $M$ is compatible with the data if the relations
\be
|m_{ee}|^{2}_{A_1, Z_1}(M) \simeq |m_{ee}|^{2}_{A_2, Z_2}(M)=...
\label{52}
\ee
are satisfied.

From (\ref{20}) it follows that for any nuclei (A,Z) the product
$$|m_{ee}|^{2}_{A, Z}(M) \,|M^{0\,\nu}(A,Z)|^{2}_{M}$$
 does not depend on the model M.
Thus, for two different models $M_1$ and $M_2$ we have
\be
|m_{ee}|^{2}_{A, Z}( M_{2})   = |m_{ee}|^{2}_{A, Z}(M_1)\,
\eta^{M_{2};M_{1}}(A, Z),
\label{53}
\ee
where
\be
\eta^{M_{2};M_{1}}(A, Z) = \frac{ |M^{0\,\nu}(A,Z)|^{2}_{M_1}} {|M^{0\,\nu}(A,Z)|^{2}_{M_{2}}}.
\label{54}
\ee
In the Table I we present the values of the coefficient $\eta(A, Z)$ for the case
of the  matrix elements calculated in \cite{nsm} (NSM) 
and  in \cite{Rodin} (the latest QPRA calculations).  

If $\eta(A, Z)$ depends on  (A, Z) and one model is compatible with the data 
the other model in principle can be excluded.
However, as it is seen from the Table I 
from the observation of the $0 \nu \beta \beta$- decay
of $^{136} \rm{Xe}$ and  $^{130} \rm{Te}$ it will be difficult to distinguish models 
\cite{nsm} and \cite{Rodin}: the difference between 
$\eta(^{136} \rm{Xe})$ and $\eta(^{130} \rm{Te})$ is about 20\% .
It will be more easier to distinguish models \cite{nsm} and \cite{Rodin}
if $0 \nu \beta \beta$- decay
of $^{76} \rm{Ge}$ and  $^{130} \rm{Te}$ is observed.

Taking into account the existence of many models of the calculations of the nuclear matrix elements 
 of the
$0 \nu \beta \beta$- decay we can conclude that the observation of 
neutrinoless double $\beta$-decay of three (or more) nuclei would  be an important 
tool in the solution of the problem of NME.

\begin{center}
                   Table I

The parameter $\eta^{NSM;QRPA}(A, Z)$, determined by
Eq. (\ref{54}), for
nuclear matrix elements of the $0 \nu \beta \beta$- decay, calculated in 
Ref.\cite{nsm} (NSM) and in Ref.\cite{Rodin} (QRPA)
\end{center}
\begin{center}

\begin{tabular}{|cc|}
\hline
Nucleus
&
 $\eta^{NSM;QRPA}$
\\
\hline
$^{76} \rm{Ge}$
&
3.1
\\
$^{130} \rm{Te}$
&
2.1
\\
$^{136} \rm{Xe}$
&
2.5
\\
\hline
\end{tabular}
\end{center}

\section{Conclusion}

After the discovery 
of  neutrino masses 
 and neutrino mixing  
the problem
of {\em the nature of neutrinos with definite masses $\nu_{i}$} 
is one of the most fundamental.
The establishment of the nature of $\nu_{i}$ 
will have a profound impact on the understanding of
the mechanism of the generation of the  neutrino masses
and mixing.

The most sensitive to the 
small Majorana neutrino masses process is neutrinoless double
$\beta$-decay of even-even nuclei. 
 From today's data to following bounds for the effective Majorana mass can be
inferred
$$|m_{ee}| \leq (0.3-1.2)\,~\rm{eV}$$
New experiments on the search for
$0\nu\beta\beta$-decay
of $^{130} \rm{Te}$, $^{76} \rm{Ge}$, $^{136} \rm{Xe}$,  $^{100} \rm{Mo}$
and other nuclei are in
preparation at present.
In these experiments
the sensitivity
$$|m_{ee}| \simeq  \rm{a\,~few}\,10^{-2}~\rm{eV}$$
is envisaged. 

The data of neutrino oscillation experiments 
allow to predict 
ranges of possible values of the effective Majorana mass
for different patterns of the
neutrino mass spectra. In order to obtain information on the 
neutrino mass spectrum
 it is important not only to observe  
$0\nu\beta\beta$-decay 
but also to {\em determine} the value of the effective Majorana mass $|m_{ee}|$.

From the measured half-life of  $0\nu\beta\beta$-decay
only the product of the effective Majorana mass and
nuclear matrix element can be obtained.
Existing calculations of the nuclear matrix elements of 
the $0\nu\beta\beta$-decay differ by about a factor of three or more.
The improvement of 
the calculations of the nuclear matrix elements is a real theoretical challenge.
We have discussed here a possible method which could allow to test models of
calculation of the nuclear matrix elements of the
$0\nu\beta\beta$-decay.
The method is based on the factorization property of the matrix element of 
$0\nu\beta\beta$-decay 
and require
observation of the $0\nu\beta\beta$-decay of several nuclei.

I would like to thank S.T. Petcov for usefull discussions. 
I acknowledge the support of  the Italien Program ``Rientro dei cervelli''.

\end{document}